\documentclass{elsart}

\usepackage{ifpdf}
\usepackage{natbib}
\usepackage{graphicx}
\usepackage{amsmath}
\usepackage{amssymb}
\usepackage{color}

\newcommand{\Swift}{{\it Swift} }
\newcommand{\swift}{\textit{Swift} }

\def\simlt{\mathrel{\hbox{\rlap{\hbox{\lower4pt\hbox{$\sim$}}}\hbox{$<$}}}}
\def\simgt{\mathrel{\hbox{\rlap{\hbox{\lower4pt\hbox{$\sim$}}}\hbox{$>$}}}}

\newcommand{\mnras}{MNRAS }
\newcommand{\apjl}{ApJ }
\newcommand{\apj}{ApJ }

\newcommand{\nat}{Nature }
\newcommand{\aap}{A\&A }

\definecolor{webgreen}{rgb}{0,.5,0}
\definecolor{webbrown}{rgb}{.6,0,0}

\ifpdf
\usepackage[%
  pdftitle={Short/Hard Gamma-Ray Burst},%
  pdfauthor={Ehud Nakar},%
  pdfsubject={},%
  pdfkeywords={},%
  pdfstartview=FitH,%
  bookmarks=true,%
  bookmarksopen=true,%
  breaklinks=true,%
  colorlinks=true,%
  linkcolor=blue,anchorcolor=blue,%
  citecolor=blue,filecolor=green,%
  menucolor=blue,pagecolor=blue,%
  urlcolor=blue]{hyperref}
\else
\usepackage[%
  breaklinks=true,%
  colorlinks=true,%
  linkcolor=blue,anchorcolor=blue,%
  citecolor=blue,filecolor=green,%
  menucolor=blue,pagecolor=blue,%
  urlcolor=blue]{hyperref}
\fi

\makeatletter
\def\elsartstyle{%
    \def\normalsize{\@setfontsize\normalsize\@xiipt{14.5}}
    \def\small{\@setfontsize\small\@xipt{13.6}}
    \let\footnotesize=\small
    \def\large{\@setfontsize\large\@xivpt{18}}
    \def\Large{\@setfontsize\Large\@xviipt{22}}
    \skip\@mpfootins = 18\p@ \@plus 2\p@
    \normalsize
} \@ifundefined{square}{}{\let\Box\square} \makeatother

\begin{document}


\begin{frontmatter}

 \title{Some Theoretical Implications of Short-Hard Gamma-Ray Burst observations}
 \author{Ehud Nakar}
 \ead{udini@tapir.caltech.edu}
 \address{Theoretical Astrophysics, California Institute of Technology, MC 130-33, Pasadena, CA 91125, USA\\
  }




\begin{abstract}
Short-hard and long-soft gamma-ray bursts (GRBs) are two distinct
phenomena, but their prompt and afterglow emission show many
similarities. This suggests that two different progenitor systems
lead to similar physical processes and that the prompt and afterglow
observations of short-hard GRBs (SHBs) can be examined using models
of long GRBs. Here, I discuss three conclusions that can be drawn
from SHB observations. I show that the lower limit on the Lorentz
factor of SHBs is typically ``only'' $10-50$, significantly lower
than that of long GRBs. SHBs with observed X-ray afterglow after 1
day are found to be roughly as efficient as long GRBs in converting
the outflow energy into prompt gamma-rays. Finally, I examine the
origin of SHBs with X-ray dark afterglows and find that the most
plausible explanation is that these SHBs exploded in extremely low
density environment ($n \lesssim 10^{-5} \rm ~cm^{-3}$).
\end{abstract}
\end{frontmatter}

\section{Introduction}
The distinct nature of short-hard gamma-ray bursts \citep[SHBs;
see][for a review]{Nakar07} was confirmed last year with the
detection of the first SHB afterglows \citep{Gehrels05,
Castro05b,Prochaska05,Fox05,Hjorth05,hwf+05,Bloom06,Covino06,Berger05}.
These observations indicate that SHBs are associated with an old
stellar population
\citep{Gehrels05,Prochaska05,Berger05,NakarGalYamFox06,Guetta06,Zheng06,Shin06}.
This is in contrast to long gamma-ray bursts (long GRBs), which are
associated with young massive star progenitors
\citep[e.g.,][]{Fruchter06} and likely produced along with a
supernova during stellar core-collapse
\citep[e.g.,][]{Stanek03,Hjorth03}. Therefore short and long GRBs
originate from distinct progenitor systems. On the other hand short
and long GRBs share many properties. There are many similarities
between the temporal structure \citep{McBreen01,Nakar02} and
spectral properties \citep{Ghirlanda04} of the prompt emission of
short and long GRBs. Observed SHB afterglows share many common
features with those observed in long GRBs \citep{Fox06,Nakar07}.
Therefore, it seems that two different progenitor systems lead to
similar physical processes and that the prompt and afterglow
observations of SHBs can be examined using models of long GRBs.

Here I discuss the interpretation of three such observations. First,
I use the observations of SHB prompt emission in order to derive the
lower limit on the Lorentz factors of the emission sources. Second,
I roughly estimate the efficiency in which gamma-rays are produced
in SHBs with observed late ($\sim 1$ d) X-ray afterglows and
finally, I discuss different explanation to SHBs with X-ray dark
afterglows. All these results and their implications are discussed
in more details in \cite{Nakar07}.

\section{The Lorentz factor of the outflow}\label{SEC: LF}
Perhaps the most prominent feature of GRBs is that they are
ultra-relativistic sources. In the case of long GRBs this is a
well-established result, relying on several independent evidence.
The main indication of high Lorentz factor in long GRBs is the
opacity constraint, where a lower limit on the Lorentz factor is set
by requiring that the source of the prompt emission is optically
thin. Some other indications that long GRB outflows propagate close
to the speed of light are resolved radio images of the afterglow of
GRB 030329 \citep{Taylor04}, scintillation quenching in the radio
afterglow of GRB 970508 \citep{Waxman98} and the onset time of the
early afterglow \citep[e.g.,][]{Sari99}. Here I use the opacity
constraint, which is the most robust and model independent method,
in order to derive a lower limit on the Lorentz factor of the source
of SHB prompt emission.

The prompt emission of short GRBs is non-thermal \citep{Lazzati05},
implying that the source is optically thin to the observed photons.
On the other hand, if a non-relativistic source is assumed, a
calculation of the optical depth, based on the enormous observed
luminosity of MeV $\gamma$-rays, results in an optical depth $\tau
\sim 10^{13}$ \citep{Schmidt78}. This conflict  can be  alleviated
if the source of the emission is moving at relativistic velocities
towards the observer \citep[e.g.,][]{Guilbert83,PiranShemi93}. The
most comprehensive calculation of the opacity limit on the Lorentz
factors of long bursts appears in \cite{Lithwick01}. Below, I carry
out similar analysis, adapting it to the prompt emission spectra of
SHBs.

\begin{figure}
\includegraphics[width=13cm]{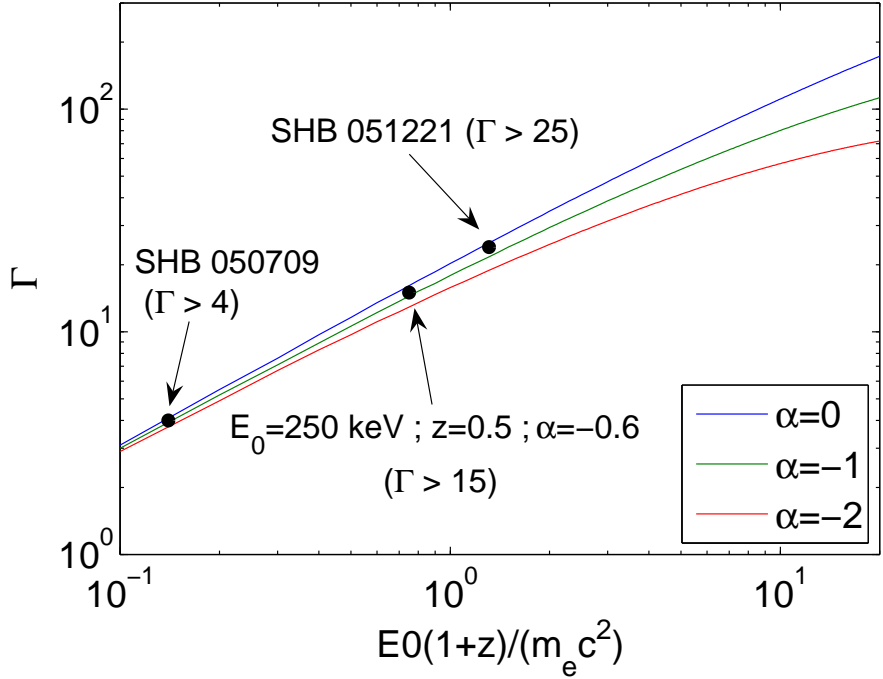}
\caption{\label{FIG: LF} {\scriptsize The lower limit on the Lorentz
factor of the prompt emission source as a function of the rest frame
spectral typical energy $E_0(1+z)$ in units of $m_ec^2$. The limit
is derived by the opacity constraint (Eq. \ref{EQ: gamma
constraint1}) using three different low energy power-law slopes
$\alpha$. The dots mark three bursts, SHB 050709, SHB 051221A
\citep[prompt emission properties are taken from
][]{Villasenor05,Golenetskii05_GCN4394}, and a typical SHB
\citep[according to][]{Ghirlanda04} at z=0.5.}
 }
\end{figure}

The non-thermal spectrum of GRBs implies that the source is
optically thin to Thompson scattering on $e^-e^+$
pairs\footnote{Opacity to $\gamma\gamma$ pair production provides
less stringent constraints on SHB Lorentz factors.}. An inevitable
source for such pairs is the annihilation of photons with rest frame
energy $\epsilon^,_{ph}
> m_ec^2$, where  $m_e$ is the electron mass. Therefore, the
Thompson optical depth for a given pulse during the prompt emission
phase is\footnote{Along the calculation I assume that the source is
moving directly toward the observer. If the source is moving at some
angle with respect to the line-of-sight then the optical depth
increases and so does the lower limit on $\Gamma$.}:
\begin{equation}\label{EQ: tau_T1}
    \tau_T \sim \frac{\sigma_T
    N_{ph} f(\epsilon^,_{ph}>m_ec^2)}{4\pi R^2},
\end{equation}
where $\sigma_T$ is the Thompson cross-section, $N_\gamma$ is the
total number of emitted photons within the pulse,
$f(\epsilon^,_{ph}>m_ec^2)$ is the fraction of photons that create
pairs and $R$ is the radius of the source. Relativistic motion of
the source has two effects. First, it reduces the rest frame energy
of the observed photons, thereby reducing $f$. Second, for a given
observed pulse time, $\delta t$, it increases the emission radius as
$R \sim c\delta t\Gamma^2$. The time scales and luminosities of
individual pulses in long and short GRBs are similar, but the
spectra may be different. While the spectrum of most long GRBs is
best described by a Band function \citep[smoothly broken
power-law;][]{Preece00,Ghirlanda02}, the best fits spectrum of most
SHBs is a low energy power-law and an exponential cut-off
\citep[PLE;][]{Ghirlanda04,Mazets04}. While this spectral difference
may be a result of observational selection
effects\footnote{Preference of a PLE fit may for example be a result
of low signal-to-noise ratio and low sensitivity of the detector at
high energies }, PLE spectrum should be used when conservatively
deriving the lowest Lorentz factor that is consistent with all
current observations. Moreover, Gev photons were observed in several
long GRBs \citep[e.g.,][]{Schneid92,Sommer94,Hurley94} while there
is no report in the literature of a photon harder than $10$ MeV that
was observed from a SHB. Using a power-law spectrum with a photon
index $\alpha$ and an exponential cut-off at $E_0$ ($dN/dE \propto
E^{\alpha} {\rm exp}\left[-E/E_0\right]$), Eq. \ref{EQ: tau_T1}
becomes:
\begin{equation}\label{EQ: tau_T2}
    \tau_T \sim 10^{14} S_{\gamma,-7} d_{L,28}^2 \delta t_{-2}^{-2}
    \frac{m_ec^2}{E_0} \Gamma^{-(4-\alpha)} {\rm exp}\left[-\frac{\Gamma m_e c^2}{E_0 (1+z)}\right],
\end{equation}
where $S_\gamma$ is the observed gamma-ray fluence of the pulse,
$d_L$ is the luminosity distance to the burst (at redshift $z$) and
throughout the paper $N_x$ denotes $N/10^{x}$ in c.g.s units.
Requiring $\tau_T<1$ results in the following constraint on the
Lorentz factor:
\begin{equation}\label{EQ: gamma constraint1}
 \frac{\Gamma m_e c^2}{E_0 (1+z)} + (4-\alpha) {\rm ln}(\Gamma)
 +{\rm ln}\left[\frac{E_0}{m_ec^2}\right] \gtrsim 30.
\end{equation}
The logarithmic dependence on $S_\gamma$, $\delta t$ and $d_L$ is
neglected in Eq. \ref{EQ: gamma constraint1} (the range of the
observed values of SHB pulses may affect the value of Eq. \ref{EQ:
gamma constraint1} by less than $50\%$). Figure \ref{FIG: LF}
presents the lower limit on $\Gamma$ as a function of $E_0$ for
three values of $\alpha$. This lower limit is $\Gamma \gtrsim 15$
for the majority of the bursts analyzed by \citet{Ghirlanda04},
assuming that they are cosmological, while for SHBs 051221 and
050709 the opacity lower limits are $\Gamma > 25$ and $\Gamma > 4$
respectively. These lower limits are significantly smaller than
those obtained for long GRBs \citep[$\Gamma \gtrsim 100$;
][]{Lithwick01}. Note however that for both populations only lower
limits on the Lorentz factor are available and, while the typical
Lorentz factor of SHBs could be significantly lower than that of
long GRBs, a precise comparison between the real values of $\Gamma$
is impossible. Additionally, the smaller lower limits on SHB Lorentz
factors depend on he best-fit function of their spectra which might
be affected by observational selection effects. Hopefully the high
energy spectra of SHBs will be securely determined by the upcoming
GLAST mission.

\section{Gamma-ray efficiency}\label{SEC: gamma eff}
\begin{figure}
\includegraphics[width=12cm]{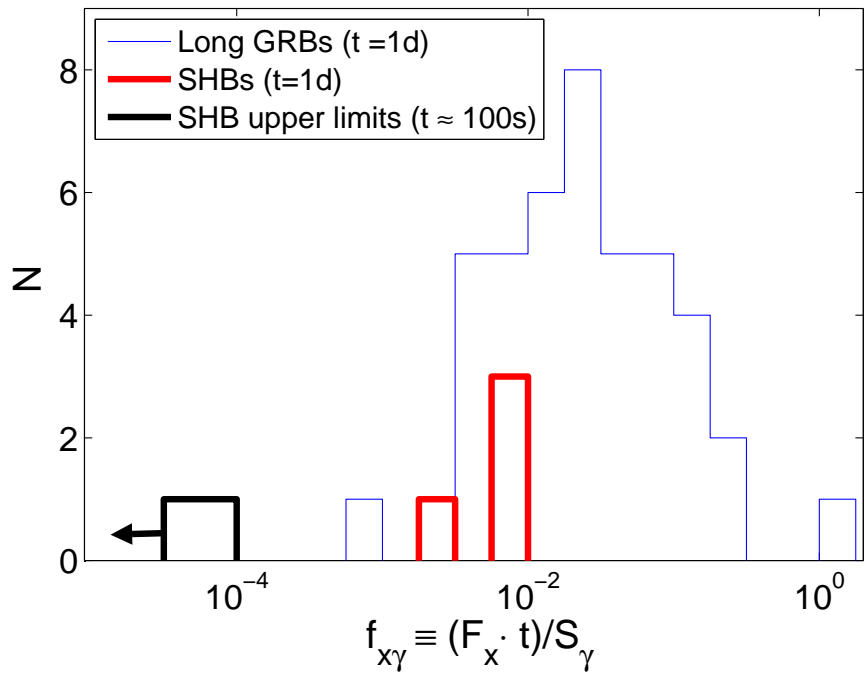}
\caption{\label{FIG: ratio_gx} {\scriptsize A histogram of the ratio
between the X-ray energy flux at time $t$ multiplied by $t$ and the
prompt gamma-ray fluence. This is an estimate of the ratio between
the energy emitted in the late X-ray afterglow and in the prompt
emission. The ratio is given for \Swift long bursts ({\it thin
line}) and for SHBs with X-ray afterglow observed after $\sim 1$ day
({\it thick line}). The upper limits at $t \sim 100$ s for two
\swift SHBs without detected X-ray afterglow are marked with arrow
({\it thick line + arrow}). Reference: \swift archive,
\href{http://swift.gsfc.nasa.gov/docs/swift/archive/grb\_table}{http://swift.gsfc.nasa.gov/docs/swift/archive/grb\_table}.
}}
\end{figure}

Figure \ref{FIG: ratio_gx} presents the distribution of the
dimensionless ratio $f_{x\gamma} \equiv F_x t/S_\gamma$ of \Swift
long and short GRBs at $t=1$ d, where $F_x$ is the X-ray ($0.2-10$
keV) energy flux at time $t$ and $S_\gamma$ is the prompt emission
gamma-ray fluence ($15-150$ keV). Within the framework of the
standard afterglow model \citep[e.g.,][]{Sari98,Granot02} and as
long as the blast wave is quasi-spherical:
\begin{equation}\label{EQ: LX_Eg_ratio}
    f_{x\gamma} \equiv \frac{F_x t}{S_\gamma} \approx \left\{
    \begin{array}{cr}
    10^{-2} \kappa^{-1} \varepsilon_{e,-1}^{3/2}  \varepsilon_{B,-2} E_{k,50}^{1/3}n^{1/2} ~~~&~~~\nu_x<\nu_c \\
    2 \cdot 10^{-3} \kappa^{-1} \varepsilon_{e,-1}^{3/2}  t_d^{-1/3} ~~~&~~~\nu_x>\nu_c
                                          \end{array}\right.
\end{equation}
where,
\begin{equation}\label{EQ: kappa}
    \kappa \equiv \frac{E_\gamma}{E_k}
\end{equation}
represents the $\gamma$-ray efficiency of the prompt emission.
$\nu_x$ is the X-ray frequency and $\nu_c$ is the synchrotron
cooling frequency. $E_k$ is the kinetic energy of the blast wave and
$E_\gamma$ is the energy emitted in $\gamma$-rays (both energies are
isotropic equivalent). Radiative loses of the blast wave energy,
which are expected to affect $E_k$ by a factor of order unity, are
neglected. The exact power of the parameters in Eq. \ref{EQ:
LX_Eg_ratio} (e.g., $\varepsilon_e$) depends weakly on $p$ (assuming
here $2.1<p<2.8$). For simplicity I use approximate power values. I
also neglect weak dependence (power-law indices below $1/4$) on
parameters, since these cannot affect the result significantly (the
lack of dependence on $n$ is exact). Following \cite{Nakar07} the
synchrotron self-Compton cooling is neglected as well.

Assuming that microphysics of collisionless shocks does not vary
significantly between bursts (either long or short) $\varepsilon_e
\approx 0.1$ and $\varepsilon_B \approx 0.01$ are adopted. Under
this assumption, for long GRBs at late time ($\sim 1$ d; but before
the jet-break), one expects $\nu_c<\nu_x$ in which case
$f_{x\gamma}$ is almost a direct measure of the $\gamma$-ray
efficiency, $\kappa$. As evident from Fig. \ref{FIG: ratio_gx}, for
long GRBs $f_{x\gamma}(1 d)=0.1-10^{-3}$ implying $\kappa_{LGRB}
\approx 0.01-1$. The high efficiency of long GRBs is a well known
result\footnote{Note that I assume here that all the energy in the
external shock at 1 d was available during the gamma-ray emission.
If this is not the case and there is a significant energy injection
into the external shock at late time, then the efficiency is even
higher
\citep[e.g.,][]{Granot06}.}\citep{Freedman01,Lloyd04,Granot06,Fan06a}.
Figure \ref{FIG: ratio_gx} also shows that the values of
$f_{x\gamma}$ for SHBs with observed X-ray afterglows are comparable
to those of long GRBs, $\sim 0.01$.

For some SHBs the circum-burst density can be low, in which case it
is not clear wether $\nu_c$ is above or below the X-ray band at 1
day. If $n \gtrsim 0.01~\rm cm^{-3}$ then $\nu_x \lesssim \nu_c$ and
$\kappa_{SHB} \sim 0.1$. If the density is significantly smaller
then $\nu_x > \nu_c$ and $\kappa$ decreases as $n^{1/2}$. Since SHBs
with observed afterglows are typically located within their host
galaxy light \citep[e.g.,][]{Fox05}, most likely $n \gg 10^{-4}~\rm
cm^{-3}$, so $\kappa \sim 0.01-0.1$. We can conclude that at least
in some SHBs (those with observed X-ray afterglow) the gamma-ray
efficiency is most likely similar to that of long GRBs
\citep[see][for a specific exploration of the efficincy of SHB
050509B]{Bloom06,Lee05}.

\section{X-ray dark afterglows}

Early X-ray afterglow from long GRBS is always detected. In
contrast, there are several SHBs with tight upper limits on any
early ($< 100$ s) X-ray emission. The values of $f_{x\gamma}
\lesssim 5 \cdot 10^{-5}$ for these bursts are exceptionally low
(Fig. \ref{FIG: ratio_gx}). Making the plausible assumptions that
the gamma-ray efficiency of these bursts is typical ($\kappa \sim
0.1$) as are the initial Lorentz factor and the microphysical
parameters, these values of $f_{x\gamma}$ indicate that these events
occurred in extremely low density environments, $n \lesssim 10^{-5}
\rm ~cm^{-3}$, typical for the inter galactic medium. This result
suggests that these SHBs occurred outside of their host galaxies.

While low density is needed to explain X-ray dark afterglows when
the most plausible assumptions are considered, alternative solutions
are viable when some of these assumptions are relaxed. For example,
assuming an inter-stellar density ($n \gtrsim 0.01 ~\rm cm^{-3}$),
the low $f_{x\gamma}$ value can be explained by ultra-efficient
gamma-rays production ($\kappa \gtrsim 100$), by unusually low
electron and magnetic field energies ($\epsilon_e^{3/2}\epsilon_B
\lesssim 10^{-6}$) or by low initial Lorentz factor $\Gamma_0
\lesssim 20$. The latter case can explain the faint early afterglow
because low $\Gamma_0$ results in a late deceleration time (and
therefore the afterglow onset), $t_{dec} \gg 100$ s.

\section{Discussion}
I considered several theoretical constraints that can be drawn from
the observations of the prompt and afterglow emission of SHBs. I
derived a constraint on the Lorentz factor of the prompt emission
source, based on the time scales, luminosity  and spectrum of the
prompt emission. These model independent lower limits imply that
SHBs are ultra-relativistic, but as $\Gamma > 10-50$ is typically
consistent with the observations they may be significantly less
relativistic than long GRBs.

Analysis of the X-ray flux of SHB afterglows after 1 day in the
context of the standard external shock model implies that SHBs are
as efficient as long GRBs in converting the energy of the
relativistic outflow to prompt gamma-rays. This result, together
with the high prompt emission variability \citep{Nakar02}, indicates
that the prompt emission is most likely a result of internal
dissipation within the outflow \citep[][; see \citealt{Nakar07} for
other indications that support this conclusion]{Sari97}.

SHBs with X-ray dark afterglows are most likely occur in a low
density environment, $n \lesssim 10^{-5} \rm ~cm^{-3}$, which is
expected in the inter-galactic medium. This result support a model
of long-lived progenitor systems ($\gtrsim 1$ Gyr) that experience a
strong natal ``kick'' ($\gtrsim 100$ km/s), as predicted in the case
of a merger of neutron star with another neutron star or a black
hole \citep[e.g.,][]{Bloom99,Belczynski06}.

\section*{Acknowledgements}
I am grateful to Avishay Gal-Yam for many helpful discussions and
comments. This work was supported by a senior research fellowship
from the Sherman Fairchild Foundation and by NASA NNH05ZDA001N
grant.


\end{document}